\documentclass[twocolumn,showpacs,showkeys]{revtex4}
\usepackage{graphicx}
\usepackage{bm}
\usepackage{color}
\usepackage{amsmath}
\usepackage{natbib}

\begin{document}

\title{NLSE for quantum plasmas with the radiation damping}

\author{Pavel A. Andreev}
\email{andreevpa@physics.msu.ru}
\affiliation{M. V. Lomonosov Moscow State University, Moscow, Russian Federation.}

\date{\today}

\begin{abstract}
We consider contribution of the radiation damping in the quantum hydrodynamic equations for spinless particles. We discuss possibility of obtaining of corresponding non-linear Schrodinger equation (NLSE) for the macroscopic wave function. We compare contribution of the radiation damping with weakly (or semi-) relativistic effects appearing in the second order by v/c. The radiation damping appears in the third order by v/c. So it might be smaller than weakly relativistic effects, but it gives damping of the Langmuir waves which can be considerable.

 \end{abstract}

% PACS, the Physics and Astronomy
                             % Classification Scheme.
%Use showkeys class option if keyword

\maketitle

%52.27.Ny	Relativistic plasmas

%52.30.Ex	Two-fluid and multi-fluid plasmas

%52.35.-g	Waves, oscillations, and instabilities in plasmas and intense beams

%%%%%%%%%%TEXT

\section{Introduction}

Some relativistic effects have been considered in classic and quantum plasmas \cite{Asenjo PRE 12}-\cite{Haas PRE 12}. In many cases one-particle Schrodinger, Pauli, Klein-Gordon, and Dirac equations have been used to derive set of quantum hydrodynamic equations \cite{Haas PRE 12}, \cite{Takabayashi}, \cite{Shukla Eliasson 2011}, \cite{Asenjo PP 11}. Some collective effects in classic relativistic plasmas with the radiation damping were considered in Ref. \cite{Kuzmenkov 80th}. A method of rigorous derivation of non-relativistic and semi-relativistic hydrodynamic equations from many-particle Schrodinger equation was suggested in Ref. \cite{Maksimov TMP 99} and developed in Refs. \cite{Ivanov Darwin}, \cite{Ivanov RPJ 13}, \cite{Andreev RPJ 07}, \cite{Andreev PRB 11}. Semi-relativistic effects appears in the second order by the parameter $v/c$ showing ratio of the particle velocity $v$ to the speed of light $c$. In this paper we are interested in contribution of the radiation damping in the evolution of quantum plasmas working in terms of quantum hydrodynamics (QHD). This effect appears in  the third order by $v/c$, when electromagnetic radiation of particles arises in theory. So system can not be described in terms of Hamiltonian of particles, the electromagnetic field have to be explicitly accounted. However the method of classic hydrodynamic derivation suggested in Refs. \cite{Drofa1996}, \cite{Kuzmenkov 91} does not apply Hamiltonian description using the Newton equations. Recent applications and discussions of this method can be found in Refs. \cite{Andreev arxiv rel} and \cite{pavelproc cl}. Hence this method gives possibility to derive hydrodynamics with the radiation damping. Comparing final equations with similar QHD equation we make generalization of obtained equation on quantum plasmas.

Hydrodynamic equations appears as natural representation of classic and quantum dynamics of many-particle systems. In some cases the set of continuity and Euler equations including the quantum Bohm potential can be represented as non-linear Schrodinger equation (NLSE) for macroscopic wave function defined in terms of the particle concentration and the potential of velocity field \cite{Maksimov TMP 99}, \cite{Andreev PRA08}. Famous examples of NLSE are the Gross-Pitaevskii equation for Bose-Einstein condensates of neutral atoms and Ginsburg-Landau equation for superconductors. Different methods for dealing with NLSEs have been developed, hence they can be used for weakly relativistic quantum plasmas with the radiation damping as well.

We gave an old example of studying of the radiation damping in plasmas \cite{Kuzmenkov 80th}. However it is a topic under discussion in recent papers as well. For instance laser-plasma interactions in ultra-relativistic regime were considered in Ref. \cite{Bashinov arxiv 13} to calculate the nonlinear dielectric
permittivity, ponderomotive and dissipative forces acting in plasmas. The motion of a cold electron fluid accounting for the radiation reaction force in the
Lorentz-Abraham-Dirac form was used (see formula 3). Some radiative and quantum electrodynamics
effects were numerically modeled for ultra-relativistic
laser-plasma interactions as well \cite{Lobet arxiv 13}. The scalar and spinor quantum electrodynamics in the
presence of strong laser fields in plasmas were considered in Ref. \cite{Raicher arxiv 13}.

\section{Construction of macroscopic equations}

Microscopic classic motion of each particle in plasmas obeys the Newton equation
\begin{equation}\label{QRD}m_{i}\dot{\textbf{v}}_{i}=\textbf{F}_{i},\end{equation}
where $i$ is the number of particle and force acting on the particle includes the radiation damping $\textbf{f}_{i}$ along with the Lorentz force
\begin{equation}\label{QRD}\textbf{F}_{i}=q_{i}\textbf{E}_{i}+\frac{q_{i}}{c}[\textbf{v}_{i},\textbf{B}_{i}]+\textbf{f}_{i},
\end{equation}
where $\textbf{E}_{i}$ and $\textbf{B}_{i}$ are electric and magnetic fields acting on $i$th particle and creating by other particles of the system, $m_{i}$ are masses of particles, $q_{i}$ are charges of particles, $\textbf{v}_{i}$ is the velocity of particles, and $c$ is the speed of light.

Non-relativistic force for the radiation damping appears as
\begin{equation}\label{QRD rad damping force}\textbf{f}_{i}\approx\frac{2q_{i}^{2}}{3c^{3}}\ddot{\textbf{v}}_{i} \approx\frac{2q_{i}^{3}}{3m_{i}c^{3}}\dot{\textbf{E}}_{i}+\frac{2q_{i}^{4}}{3m_{i}^{2}c^{4}}[\textbf{E}_{i},\textbf{B}_{i}]
\end{equation}
(see Ref. \cite{Landau 2} section 9, paragraph 75, formula 75.8). The second identity in formula (\ref{QRD rad damping force}) has been made taking derivative of the acceleration with respect to time and neglecting $\dot{\textbf{f}}$. In the second term of the Lorentz force we have used $\dot{\textbf{v}}=e\textbf{E}$, and we have not considered time derivative of the magnetic field. Doing these approximations we keep considering terms in the third order on $v/c$.

As the result of manipulations described above we have an approximate equations for classic motion of electrons with the radiation damping. We can use it as framework to derive classic hydrodynamic equations describing collective evolution of considering system. To this end we use method suggested in Ref. \cite{Drofa1996} and briefly reviewed in Ref. \cite{pavelproc cl}. This method gives the following continuity and Euler equations
\begin{equation}\label{QRD cont eq}\partial_{t}n+\nabla (n\textbf{v})=0
\end{equation}
and
$$mn(\partial_{t}+\textbf{v}\nabla)\textbf{v}+\nabla p=qn\textbf{E}$$
\begin{equation}\label{QRD Euler Cl} +\frac{q}{c}n[\textbf{v},\textbf{B}]+\frac{2q^{3}}{3mc^{3}}\dot{\textbf{E}}+\frac{2q^{4}}{3m^{2}c^{4}}[\textbf{E},\textbf{B}]
\end{equation}
written in the self-consistent field approximation allowing to truncate the chain of HD equations at using of an equation of state for the thermal pressure. In equations (\ref{QRD cont eq}) and (\ref{QRD Euler Cl}) we have next physical quantities $n$ is the particle concentration, $\textbf{v}$ is the velocity field, $\partial_{t}$ and $\nabla$ are the time and spatial derivatives, and $p$ is the thermal pressure. Different terms in equations (\ref{QRD cont eq}) and (\ref{QRD Euler Cl}) have following meaning. The continuity equation (\ref{QRD cont eq}) shows conservation of the particles number and gives time evolution of the particle concentration. The Euler equation (\ref{QRD Euler Cl}) is the equation of particles current (particle flux) evolution. In non-relativistic systems it coincides with the equation for the momentum current (or the momentum flux). In fact we should present all weakly relativistic terms in equation (\ref{QRD Euler Cl}), but we do not want to present rather huge equation, for the weakly relativistic effects in quantum hydrodynamics see Refs. \cite{Ivanov Darwin}, \cite{Ivanov RPJ 13}. These equations are coupled with the equations of field
\begin{equation}\label{QRD divE}\nabla \textbf{E}(\textbf{r},t)=4\pi \sum_{a}q_{a}n_{a}(\textbf{r},t),
\end{equation}
\begin{equation}\label{QRD curlE}\nabla\times \textbf{E}(\textbf{r},t)=0,
\end{equation}
\begin{equation}\label{QRD divE}\nabla \textbf{B}(\textbf{r},t)=0,
\end{equation}
and
\begin{equation}\label{QRD curlE}\nabla\times \textbf{B}(\textbf{r},t)=\frac{4\pi}{c}\sum_{a}q_{a}n_{a}(\textbf{r},t)\textbf{v}_{a}(\textbf{r},t).
\end{equation}
We do not have time derivatives of field here, since only trace of radiation in the approximation under consideration is the radiation damping.

We see a force field caused by the radiation damping. Comparing classic and quantum hydrodynamic equations we can get contribution of the radiation damping in the set of QHD equations. The quantum Euler equation differs from the classic one by the quantum Bohm potential and has following form
$$mn(\partial_{t}+\textbf{v}\nabla)\textbf{v}+\nabla p$$
$$-\frac{\hbar^{2}}{2m}\nabla\Biggl(\frac{\triangle n}{n}-\frac{(\nabla n)^{2}}{2n^{2}}\Biggr)=qn\textbf{E}$$
\begin{equation}\label{QRD Euler Quant} +\frac{q}{c}n[\textbf{v},\textbf{B}]+\frac{2q^{3}}{3mc^{3}}\dot{\textbf{E}}+\frac{2q^{4}}{3m^{2}c^{4}}[\textbf{E},\textbf{B}].
\end{equation}
Considering quantum plasmas of charged Fermi particles we should consider the Fermi pressure instead of thermal pressure.

Having equations (\ref{QRD cont eq}) and (\ref{QRD Euler Quant}) we can derive a non-linear Schrodinger equation for the macroscopic wave function \cite{Andreev PRB 11} neglecting by the magnetic field and assuming that the electric field is potential $\textbf{E}=-\nabla\varphi$
\begin{equation}\label{QRD wave func macro definition}\Phi=\sqrt{n}\exp\biggl(\frac{\imath}{\hbar}m\phi\biggr),
\end{equation}
where $\phi$ is the potential of the velocity field $\textbf{v}=\nabla\phi$.

Considering the potential electric field we can introduce an effective potential electric field including contribution of the radiation damping
\begin{equation}\label{QRD}\widetilde{\textbf{E}}=-\nabla\widetilde{\varphi}=-\biggl(1+\frac{2e^{2}}{3mc^{3}}\partial_{t}\biggr)\nabla\varphi.
\end{equation}
Differentiating the macroscopic wave function (\ref{QRD wave func macro definition}) with respect to time we obtain a non-linear Schrodinger equation
\begin{equation}\label{QRD}\imath\hbar\partial_{t}\Phi(\textbf{r},t)
=\biggl(-\frac{\hbar^{2}\nabla^{2}}{2m}+(3\pi^{2})^{\frac{2}{3}}\frac{\hbar^{2}}{2m}n^{\frac{2}{3}}(\textbf{r},t)+q\widetilde{\varphi}\biggr)\Phi(\textbf{r},t) \end{equation}
with $n=\mid\Phi\mid^{2}$.

Writing explicit form of the potential of effective electric field we represent the NLSE as follows
$$\imath\hbar\partial_{t}\Phi(\textbf{r},t)
=\biggl(-\frac{\hbar^{2}\nabla^{2}}{2m}$$
\begin{equation}\label{QRD} +(3\pi^{2})^{\frac{2}{3}}\frac{\hbar^{2}}{2m}n^{\frac{2}{3}}(\textbf{r},t)+q\varphi+\frac{2q^{3}}{3mc^{3}}\partial_{t}\varphi\biggr)\Phi(\textbf{r},t), \end{equation}
where the potential of electric field depends on the particle concentration via the Maxwell equations (\ref{QRD divE}), (\ref{QRD curlE}), and consequently it depends on the macroscopic wave function. Hence we have closed set of equations consisting of the NLSE and the Maxwell equations.

The dispersion dependence of semi-relativistic Langmuir waves in absence of the radiation damping was obtained in Ref. \cite{Ivanov Darwin}. It has the following form
$$\omega^2_{SR}(k)=\omega_{Le}^2\biggl(1-\varsigma\frac{\hbar^2k^2}{4m^2c^2}-\frac{5T}{2mc^{2}}\biggr)$$
\begin{equation}\label{QRD semi relativ spectrum}+
\frac{\hbar^2k^4}{4m^2}-\frac{\hbar^4k^6}{8m^4c^2}+\frac{\gamma T_{0}}{m}k^2,
\end{equation}
where we included shifts of the Langmuir frequency by both the semi-relativistic part of kinetic energy and the Darwin interaction presented by term proportional to $\varsigma$. Simultaneous account of both effects gives $\varsigma=0$, for details see Ref. \cite{Ivanov Darwin}. The last part of the first term appears as consequence of the thermal\textbf{-}semi-relativistic force field. Other terms have following meaning: the quantum Bohm potential, the semi-relativistic part of the quantum Bohm potential, and the contribution of the thermal pressure.

In linear approximation on small perturbations of hydrodynamic variables $\delta n=N\exp(-i\omega t+i\textbf{k}\textbf{r})$ and $\delta \textbf{v}=\textbf{U}\exp(-i\omega t+i\textbf{k}\textbf{r})$ we can get a dispersion equation giving $\omega(k)$
\begin{equation}\label{QRD disp eq with damp}\omega^{2}+\imath\lambda\omega-\Omega^{2}=0.
\end{equation}
Dispersion dependence $\omega(k)$ arises as
\begin{equation}\label{QRD disp gen with damp}\omega=\frac{1}{2}\biggl(-\imath\lambda\pm2\Omega\sqrt{1-\frac{\lambda^{2}}{4\Omega^{2}}}\biggr),
\end{equation}
including weakly-relativistic nature of the radiation damping we can give approximate formula for the dispersion dependence
\begin{equation}\label{QRD disp approx with damp}\omega\approx\Omega\biggl[1-\frac{\lambda^{2}}{8\Omega^{2}}-\imath\frac{\lambda}{2\Omega}\biggr]
\end{equation}
containing
\begin{equation}\label{QRD lambda definition}\lambda=\frac{8\pi e^{4}n_{0}}{3m^{2}c^{3}}=\frac{2}{3}\frac{\omega_{Le}^{2}r_{e}}{c},
\end{equation}
where $r_{e}\equiv e^{2}/(mc^{2})$ is the classic radius of electron, and $\omega_{Le}$ is the Langmuir frequency, \emph{and}
\begin{equation}\label{QRD Omega definition}\Omega^{2}=\frac{1}{m}\frac{\partial p}{\partial n}k^{2}+\frac{\hbar^{2}k^{4}}{4m^{2}}+\omega_{Le}^{2}
\end{equation}
with $\frac{\partial p}{\partial n}=\gamma T_{0}$, and $\gamma=3$.

In formula (\ref{QRD disp approx with damp}) we can see imaginary part of the frequency, which is negative and gives to damping of the semi-relativistic quantum Langmuir waves.

Presence of the semi-relativistic effects given by formula (\ref{QRD semi relativ spectrum}) changes $\Omega$, but it does not affect structure of solution (\ref{QRD disp gen with damp}).

Proper consideration based on full spinless semi-relativistic theory \cite{Ivanov Darwin} gives $\omega^2_{SR}(k)$ (\ref{QRD semi relativ spectrum}) instead of $\Omega$. Finally we can rewrite formula (\ref{QRD disp approx with damp}) as follows
\begin{equation}\label{QRD omega expl Re} Re\omega=\omega_{SR}(k)-\frac{1}{18}\frac{r_{e}^{2}}{c^{2}} \frac{\omega_{Le}^{4}}{\sqrt{\omega_{Le}^2+\frac{3 T_{0}}{m}k^2+\frac{\hbar^{2}k^{4}}{4m^{2}}}},\end{equation}
and
\begin{equation}\label{QRD omega expl Im} Im\omega=-\frac{1}{3}\omega_{Le}^{2}\frac{r_{e}}{c}.\end{equation}
Formula shows that the damping of Langmuir waves caused by the radiation damping do not contain contribution of quantum effects.

%\begin{equation}\label{QRD} \end{equation}

\section{Conclusion}

We have considered weakly relativistic evolution of quantum plasmas including radiation damping. We have derived corresponding equations of quantum hydrodynamics. Solving these equations we obtained spectrum of the Langmuir waves. We have found that the radiation damping gives a shift of the real part of frequency of the Langmuir waves, but it also leads to damping of the waves. We should mention that this damping is not affected by the quantum effects in considered approximation.

We have shown that the QHD equations with the radiation damping can be represented in form of a NLSE for the wave function in medium. This equation may get its place in list of the Ginsburg-Landau and Gross-Pitaevskii equations. Each of them has been very useful in their own fields. Therefore the NLSE derived in this paper opens similar possibilities for quantum plasmas.

Obtained in this paper equations allows to study different effects in the weakly relativistic quantum plasmas with the radiation damping.

\end{document}